# Radar Signal Delay in the Dvali-Gabadadze-Porrati Gravity in the Vicinity of the Sun

Ioannis Haranas[1], Omiros Ragos[2] Ioannis Gkigkitzis[3]

[1] *Department of Physics and Astronomy, York University, 4700 Keele Street, Toronto, Ontario, M3J 1P3, Canada*
*e:mail: yiannis.haranas@gmail.com*

[2]*Dept. of Mathematics, University of Patras, GR-26500 Patras, Greece*
e-mail: ragos@math.upatras.gr

[3]*Department of Mathematics, East Carolina University, 124 Austin Building, East Fifth Street Greenville, NC 27858-4353, USA*
e-mail: gkigkitzisi@ecu.edu

**Abstract**
In this paper we examine the recently introduced Dvali-Gabadadze-Porrati (DGP) gravity model. We use the space time metric in which the local gravitation source dominates the metric over the contributions from the cosmological flow. Anticipating ideal possible solar system effects we derive expressions for the signal time delays in the vicinity of the sun, and for various angles of the signal approach. We use the corresponding numerical value for the parameter $r_0$ to be equal to 5 Mpc, and from that we calculate that the time contribution due to DGP correction to the metric is proportional to $b^{3/2}/c^2 r_0$. In the vicinity of the Sun and with $\theta$ in the range $-\pi/3 \leq \theta \leq -\pi/3$, $\Delta t$ is equal to 0.0001233 ps. A time signal delay extremely small to measure by today's technology could be probably measurable in the future years to come, by various future experiments.

## 1. Introduction
There is recent attention for a five dimensional gravity model the so called Dvali-Gabadadze-Porrati (DGP). This model explains the observed acceleration of the expansion of the Universe. Furthermore, it predicts minor post-Einstein effects, testable at local scales resulting to information on the Universe's global properties in relation to the ongoing cosmological expansion (Iorio, 2005). So far, two-body scenarios have been investigated in which the time rates of change for the longitude of pericenter and the mean anomaly of the secondary have been carried out (Lue and Starkman, 2003, Iorio, 2000b). These effects are functions of eccentricity up to $O(e^2)$ but they do not depend on the semimajor axis *a* of the secondary. Following Iorio (2000b, 2000c) one might say that the "ideal test-bed for such tests is the inner planets of the solar system". Measurements of such precessions lie in the limit of precision of today's planetary data. For a more detailed and complete overview on the DGP gravity see Lue, (2003).

The DGP model is based on an extra flat dimension, and a free crossover parameter $r_0$ which defines a radius beyond which the four-dimensional gravitational theory transitions into a five-dimensional regime, and is defined as follows $r_0 = k^2/2\mu$. The constants $\mu^2$ and $k^2$ define the energy scales of the theories of gravity: one is Newton's constant, $\mu^2 = 8\pi G$, the other represents the energy scale of the bulk gravity (Sawicki, et al., 2007). The crossover parameter is also fixed from observations of IA type supernova to a value approximately equal to $\approx$ 5 Gpc (Lue and Starkman 2003). For distances greater than that Newtonian gravity needs to be modified, resulting thus to different explanations from that of dark matter in order to interpret the accelerations observed in the Universe.

In this short contribution we will calculate the signal time delay in the vicinity of a massive body of mass *M*, i.e. the Sun when the Newtonian-Einstein gravity is mod0ified due to the Dvali-Gabadadze-Porrati braneworld model. We proceed with in a way similar to Haranas and Ragos (2011) and Haranas, et al. (2011). We base our analysis on the angle $\theta$ defined by the distance of the closest approach, and any other point on the traveling radar signal. We apply our results in the vicinity of the Sun and for various angular ranges.

## 2. The Dvali-Gabadadze-Porrati Metric

Following Lue and Starkman (2003) the metric of a spherical source in a cosmological de Sitter background can be described by the following line element:

$$ds^2 = c^2 N^2(r,w)dt^2 - A^2(r,w)dr^2 - B^2(r,w)(d\theta^2 + \sin^2 d\varphi^2) - dw^2. \qquad (1)$$

In the region where the local gravitation dominates the metric over the contributions from the cosmological flow, the de Sitter solution of the five-dimensional field equations can be expressed by (ibid, 2003):

$$N(r,z) = 1 + n(r,z) = 1 - \frac{GM}{rc^2} \pm \sqrt{\frac{GMr}{r_0^2}} \qquad (2)$$

and

$$A(r, z) = 1 + \alpha(r, z) = 1 + \frac{GM}{rc^2} \mp \sqrt{\frac{GMr}{r_0^2}} \tag{3}$$

$$B(r, z) = r(1 + b(r, z)) \tag{4}$$

where *M* is gravitating mass, *G* is the constant of universal gravitation, the *w* is the fourth special coordinate, $b(r, z)$ are functions of coordinates and time, that can be calculated using the derived field equations. For distances scales much smaller than $r \ll r_0$ Newton-Einstein gravity is obtained with few exceptions that include minor corrections. With reference to Lue and Starkman (2003) wee can write the Dvali-Gabadadze-Porrati (DGP) line element in the following way:

$$ds^2 = c^2\left(1 - \frac{GM}{rc^2} \pm \sqrt{\frac{GMr}{r_0^2}}\right)^2 dt^2 - \left(1 + \frac{GM}{rc^2} \mp \sqrt{\frac{GMr}{r_0^2}}\right)^2 dr^2 - r^2(1 + b(r, z))^2 d\Omega^2 - dw^2. \tag{5}$$

To deal with the Dvali-Gabadadze-Porrati (DGP) effect on the propagation of electromagnetic signals in the vicinity of the Sun, we incorporate an additional DGP correction term in the Schwarzschild space-time metric coefficients of the line element. We analogously modify the Schwarzschild metric used in the solar system when general relativistic effects are taken into account. Therefore, if $r, \theta, \phi$ are the polar coordinates of any point along the signal's path, and $\Omega$ is the corresponding solid angle, the line element takes the form. Next, the photon transmission time can be written as follows:

$$dt' = \frac{ds}{c}, \tag{6}$$

and

$$dt = \left(1 + \frac{GM}{rc^2} \mp \sqrt{\frac{GMr}{r_0^2}}\right)^2 dt' = \left(1 + \frac{GM}{rc^2} \mp \sqrt{\frac{GMr}{r_0^2}}\right)^2 \frac{ds}{c}, \tag{7}$$

where the plus sign is related to a the Friedman-Lemaitre-Robertson-Walker phase of the universe, and where the minus sign is related to a self accelerating phase (Iorio, 2005a). In this paper we first consider the plus sign or Friedman-Lemaitre-Robertson-Walker phase first and second the minus sign representing the accelerating phase. Next, using $r = b/\cos\theta$, where *b* is the distance of the closest signal approach, and therefore we have that:

$$\mathrm{d}r = -\frac{b\sin\theta}{\cos^2\theta}\mathrm{d}\theta, \tag{8}$$

Also, $\mathrm{d}r = \sqrt{\mathrm{d}r^2 + r^2\mathrm{d}\theta^2}$ and after substitution and simplification, Eq (5) becomes:

$$\mathrm{d}t = \frac{b}{c}\left(\begin{array}{l} 1 + \frac{2GM}{bc^2}\cos\theta + \frac{G^2M^2}{b^2c^4}\cos^2\theta \\ + \frac{GMb}{c^2r_0^2}\sec\theta + 2\sqrt{\frac{GMb}{c^2r_0^2}}\sec\theta + \frac{2GM}{bc^2}\sqrt{\frac{GMb\sec\theta}{c^2r_0^2}} \end{array}\right)\sec^2\theta\,\mathrm{d}\theta. \tag{9}$$

Equation (9) contains classical, general relativistic, DGP time delays. Since we are interested in the DGP delay only we may neglect the terms $1 + \frac{2GM}{bc^2}\cos\theta + \frac{G^2M^2}{b^2c^4}\cos^2\theta$ we then have to integrate the following expression:

$$\mathrm{d}t = \frac{b}{c}\left(\frac{GMb}{c^2r_0^2}\sec\theta + 2\sqrt{\frac{GMb}{c^2r_0^2}}\sec\theta + \frac{2GM}{bc^2}\sqrt{\frac{GMb\sec\theta}{c^2r_0^2}}\right)\sec^2\theta\,\mathrm{d}\theta, \tag{10}$$

Omitting order $O(c^{-3})$ and $O(c^{-4})$ terms for being to small, we integrate over various angular subintervals of the range $(-\pi/2, \pi/2)$ to avoid the singularities at $\theta = \pm\pi/2$. For any such interval $[a,b]$ the corresponding radar signal time delay will be:

$$\Delta t = \frac{b}{c}\int_{-\pi/6}^{+\pi/6}\left(\frac{GMb}{c^2r_0^2}\sec\theta + \frac{2}{cr_0}\sqrt{GMb\sec\theta}\right)\sec^2\theta\,\mathrm{d}\theta, \tag{11}$$

We start at $[-\pi/6, \pi/6]$ and integrating we obtain that:

$$\Delta t = \frac{2GMb^2}{3c^3r_0^2} + \frac{GMb^2\ln(27)}{6c^3r_0^2} + \frac{8b\sqrt{2bGM}}{3^{7/4}c^2r_0} + \frac{8b\sqrt{bGM}}{3c^2r_0}F\left(\frac{\pi}{12}, 2\right) \tag{12}$$

where, $F$ is the elliptic integral function of the first kind (Spiegel, 1968). Similarly, integrating over the range $[-\pi/4, \pi/4]$, we obtain that:

$$\Delta t = \frac{GMb^2\sqrt{2}}{c^3r_0^2} + \frac{8\times 2^{1/4}b\sqrt{GMb}}{3c^2r_0} + \frac{8b\sqrt{GMb}}{3c^2r_0}F\left(\frac{\pi}{8}, 2\right) + \frac{GMb^2}{c^3r_0^2}\ln\left[\frac{\cos(\pi/8)+\sin(\pi/8)}{\cos(\pi/8)-\sin(\pi/8)}\right]. \tag{13}$$

Next in the range $[-\pi/3, \pi/3]$ we obtain that:

$$\Delta t = \frac{b^2\sqrt{2GM}}{c^3 r^2} + \frac{8b}{c^2 r_0}\sqrt{\frac{2GMb}{3}} + \frac{8b\sqrt{GMb}}{3c^2 r}F\left(\frac{\pi}{6},2\right) + \frac{GMb^2}{c^3 r_0^2}\ln\left(\frac{1+\sqrt{3}}{1-\sqrt{3}}\right), \quad (14)$$

finally in the range [-4π/10, 4π/10]or ± 72° we obtain the following expression

$$\Delta t = \frac{GMb^2\sqrt{50+22\sqrt{5}}}{c^3 r_0^2} + \frac{8b\sqrt{bGM}\sqrt{15+7\sqrt{5}}}{3c^2 r_0} + \frac{8b\sqrt{GMb}}{3c^2 r_0}F\left(\frac{\pi}{5},2\right) + \frac{GMb^2}{c^3 r_0^2}\ln\left(\frac{1+\sqrt{5}+\sqrt{10-2\sqrt{5}}}{1+\sqrt{5}-\sqrt{10-2\sqrt{5}}}\right).$$

(15)

Next, in the same way, we proceed to calculate the signal delays when the DGP correction to the metric appears with a negative sign. For any interval $[a,b]\subset(-\pi/2,\pi/2)$ the corresponding radar signal time delay will be:

$$\Delta t = \frac{b}{c}\int_{-\pi/6}^{+\pi/6}\left(\frac{GMb}{c^2 r_0^2}\sec\theta - \frac{2}{cr_0^2}\sqrt{GMb\sec\theta}\right)\sec^2\theta\, d\theta. \quad (16)$$

For $[-\pi/6,\pi/6]$, we obtain:

$$\Delta t = \frac{2GMb^2}{3c^3 r_0^2} - \frac{8b\sqrt{2GMb}}{3^{7/4}c^2 r_0} - \frac{8b\sqrt{GMb}}{3c^2 r_0}F\left(\frac{\pi}{12},2\right) + \frac{GMb^2}{6c^3 r_0^2}\ln(27), \quad (17)$$

where $F$ is the elliptic function of the first kind. Similarly, and for the same angular limits as before for $[-\pi/4,\pi/4]$, we obtain:

$$\Delta t = \frac{\sqrt{2}GMb^2}{c^3 r_0^2} - \frac{8\times 2^{1/4}b\sqrt{GMb}}{3c^2 r_0} - \frac{8b\sqrt{GMb}}{3c^2 r_0}F\left(\frac{\pi}{8},2\right) + \frac{GMb^2}{c^3 r_0^2}\ln\left(\frac{\cos(\pi/8)+\sin(\pi/8)}{\cos(\pi/8)-\sin(\pi/8)}\right), \quad (18)$$

next, for $[-\pi/3,\pi/3]$, we obtain:

$$\Delta t = \frac{GMb^2}{c^3 r_0^2} - \frac{8b}{c^2 r_0}\sqrt{\frac{2GMb}{3}} - \frac{8b\sqrt{GMb}}{3c^2 r_0}F\left(\frac{\pi}{6},2\right) + \frac{GMb^2}{c^3 r_0^2}\ln\left[\frac{1+\sqrt{3}}{-1+\sqrt{3}}\right] \quad (19)$$

and finally, for -4π/10< $\theta$ < 4π/10 we obtain that:

$$\Delta t = \frac{\sqrt{50+22\sqrt{5}}GMb^2}{c^3 r_0^2} - \frac{8b\sqrt{15+7\sqrt{5}}\sqrt{GMb}}{3c^2 r_0} - \frac{8b\sqrt{GMb}}{3c^2 r_0}F\left(\frac{\pi}{5},2\right) + \frac{GMb^2}{c^3 r_0^2}\ln\left[\frac{1+\sqrt{5}+\sqrt{10-2\sqrt{5}}}{1+\sqrt{5}-\sqrt{10-2\sqrt{5}}}\right]. \quad (20)$$

In the numerical results section, the evaluation of the elliptic function $F$ will be necessary. This can be accomplished by using the following up to 4$^{\text{th}}$ order series expansion of $F$ given below:

$$F(\theta,k) \approx \theta + \left(\frac{\sin^{-1}(\sin\theta)-\sin\theta\sin^2\theta}{4}\right)k + \left(\begin{array}{c}\dfrac{9(\sin^{-1}(\sin\theta)-\cos^2\theta\sin\theta)}{64} \\ -\dfrac{3\cos^2\theta\sin^3\theta}{32}\end{array}\right)k^2$$

$$+ \left(\begin{array}{c}\dfrac{25(\sin^{-1}(\sin\theta)-\sin\theta\sin^2\theta)}{256} \\ -\dfrac{25\cos^2\theta\sin^3\theta}{384} - \dfrac{5\cos^2\theta\sin^5\theta}{96}\end{array}\right)k^3 \quad (21)$$

$$+ \left(\begin{array}{c}\dfrac{1225(\sin^{-1}(\sin\theta)-\cos^2\theta\sin\theta)}{16384} - \dfrac{1225\cos^2\theta\sin^3\theta}{24576} \\ -\dfrac{245\cos^2\theta\sin^5\theta}{6144} - \dfrac{35}{1024}\cos^2\theta\sin^7\theta\end{array}\right)k^4 + O(k^5).,$$

whose numerical values are going to be substituted for the corresponding functions in Eqs. (12) to (20) in the section of the numerical results. Therefore we find that, the evaluation to $O(k^4)$ of the elliptic integrals of the first time we obtain:

$$F\left(\frac{\pi}{12},2\right) \approx -\frac{585}{512} + \frac{3339\sqrt{3}}{16384} + \frac{1379\pi}{4096} \approx 0.268087, \quad (22)$$

$$F\left(\frac{\pi}{8},2\right) \approx \frac{413}{1024} - \frac{285}{128\sqrt{2}} + \frac{4137\pi}{8192} \approx 0.415422, \quad (23)$$

$$F\left(\frac{\pi}{6},2\right) \approx -\frac{14491\sqrt{3}}{16384} + \frac{1379\pi}{2048} \approx 0.583429, \quad (24)$$

$$F\left(\frac{\pi}{5},2\right) \approx -\frac{\sqrt{3309806810-6599649875}}{32768} + \frac{4137\pi}{5120} \approx 0.744014. \quad (25)$$

## 3. Numerical Results

To apply the above analysis in the case of the Sun, we have used that we use the numerical values for the mass of the Sun $M_S = 1.99 \times 10^{30}$ kg, and taking $b \approx R_S = 6.96 \times 10^8$ m, and using $r_0 \approx 5$ Gpc $= 1.542 \times 10^{23}$ m (Lue and Starkman 2003), we tabulate in table 1 the following time signal delays in the vicinity of the Sun for the angular ranges indicated. First, the signal time delay in the vicinity of the Sun and in the range $-\pi/6 \leq \theta \leq \pi/6$ for a Friedman-Lemaitre-Robertson-Walker phase first, we have:

$$\Delta t = 5.977 \times 10^{-6} \left(\frac{b^2}{r_0^2}\right) + 3.033 \times 10^{-7} \left(\frac{b^{3/2}}{r_0}\right) \approx 3.033 \times 10^{-7} \left(\frac{b^{3/2}}{r_0}\right), \tag{26}$$

similarly, in the range $-\pi/4 \leq \theta \leq \pi/4$ we obtain:

$$\Delta t = 1.128 \times 10^{-5} \left(\frac{b^2}{r_0^2}\right) + 5.477 \times 10^{-7} \left(\frac{b^{3/2}}{r_0}\right) \approx 5.477 \times 10^{-7} \left(\frac{b^{3/2}}{r_0}\right). \tag{27}$$

Next, in the range $-\pi/3 \leq \theta \leq \pi/3$ we obtain

$$\Delta t = 2.350 \times 10^{-5} \left(\frac{b^2}{r_0^2}\right) + 1.035 \times 10^{-6} \left(\frac{b^{3/2}}{r_0}\right) \approx 1.035 \times 10^{-6} \left(\frac{b^{3/2}}{r_0}\right). \tag{28}$$

Finally, in the range $-4\pi/10 \leq \theta \leq 4\pi/10$ we obtain:

$$\Delta t = 5.802 \times 10^{-5} \left(\frac{b^2}{r_0^2}\right) + 2.144 \times 10^{-6} \left(\frac{b^{3/2}}{r_0}\right) \approx 2.144 \times 10^{-6} \left(\frac{b^{3/2}}{r_0}\right), \tag{28}$$

where the time is given in picoseconds [ps], is tabulated in table 1 below:

**Table 1**. Radar signal time delays due to the DGP gravity in the vicinity of the Sun. Results related to the Friedman-Lemaitre-Robertson-Walker phase of the universe.

| Angular Range of Closest Approach [rad] | Signal Time Delays [ps] |
|---|---|
| $-\pi/6 \leq \theta \leq \pi/6$ | 0.0000361 |
| $-\pi/4 \leq \theta \leq \pi/4$ | 0.0000652 |
| $-\pi/3 \leq \theta \leq \pi/3$ | 0.0001233 |
| $-4\pi/10 \leq \theta \leq 4\pi/10$ | 0.0002553 |

Similarly, for the accelerating phase of the universe and for the same order of angles we obtain that:

$$\Delta t = 5.977 \times 10^{-6} \left(\frac{b^2}{r_0^2}\right) - 3.033 \times 10^{-7} \left(\frac{b^{3/2}}{r_0}\right) \approx -3.033 \times 10^{-7} \left(\frac{b^{3/2}}{r_0}\right), \tag{29}$$

$$\Delta t = 1.128 \times 10^{-5} \left(\frac{b^2}{r_0^2}\right) - 5.477 \times 10^{-7} \left(\frac{b^{3/2}}{r_0}\right) \approx -5.477 \times 10^{-7} \left(\frac{b^{3/2}}{r_0}\right), \tag{30}$$

$$\Delta t = 2.350 \times 10^{-5} \left(\frac{b^2}{r_0^2}\right) - 1.035 \times 10^{-6} \left(\frac{b^{3/2}}{r_0}\right) \approx -1.035 \times 10^{-6} \left(\frac{b^{3/2}}{r_0}\right), \tag{31}$$

$$\Delta t = 5.802 \times 10^{-5} \left(\frac{b^2}{r_0^2}\right) - 2.144 \times 10^{-6} \left(\frac{b^{3/2}}{r_0}\right) \approx -2.144 \times 10^{-6} \left(\frac{b^{3/2}}{r_0}\right), \tag{32}$$

where the time is given in picoseconds [ps], is tabulated in table 2 below

**Table 2** Radar signal time delays due to the DGP gravity in the vicinity of the Sun. Results related to the accelerating phase of the universe.

| Angular Range of Closest Approach [rad] | Signal Time Delays [ps] |
|---|---|
| $-\pi/6 \leq \theta \leq \pi/6$ | -0.0000361 |
| $-\pi/4 \leq \theta \leq \pi/4$ | -0.0000652 |
| $-\pi/3 \leq \theta \leq \pi/3$ | -0.0001233 |
| $-4\pi/10 \leq \theta \leq 4\pi/10$ | -0.0002553 |

With reference to Haranas and Ragos (2011) we say that to get an idea of today's radar systems, somebody could talk about the sensitivity of a radar, a property that is related to the power of the transmitting radar. Since we are interested in signal time delays and in order to substantiate our finding we will referrer to today's radar resolution instead something that is related to the detectable times. Quoting Shapiro (1968, 1999), we say that fractional system errors of echo time delays in solar system experiments can be up to 1 part in $10^{10}$ or smaller. Given today's technological progress it might be possible that such effects will be detected in the years to come.

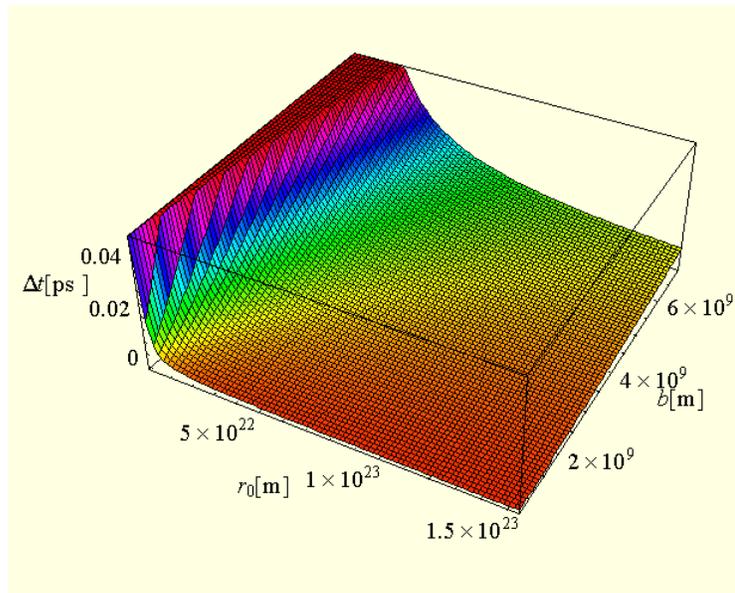

**Fig. 1** Signal time delay in the vicinity of the Sun as a function of the DGP parameter $r_0$ and distance of closest approach $b$ in Friedman-Lemaitre-Robertson-Walker phase of the universe, in the range $[-4\pi/10, 4\pi/10]$.

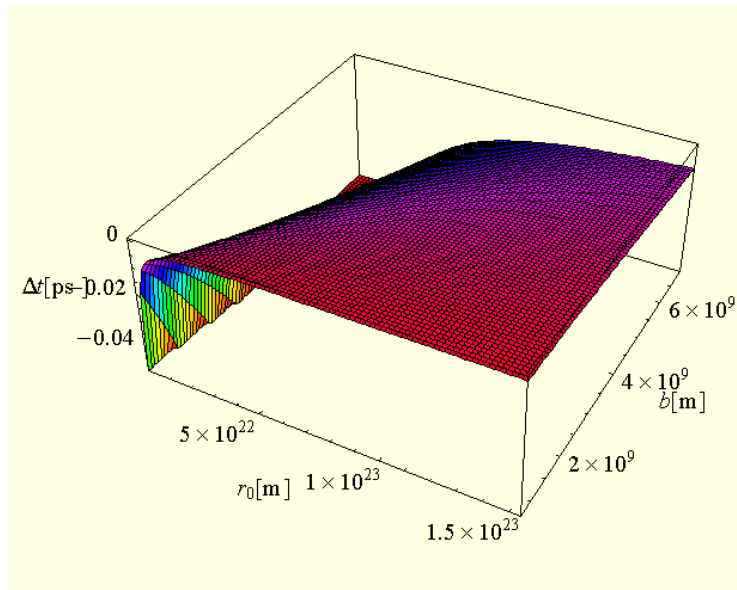

**Fig. 1** Signal time delay in the vicinity of the Sun as a function of the DGP parameter $r_0$ and distance of closest approach $b$ in related to the accelerating phase of the universe in the range $[-4\pi/10, 4\pi/10]$.

## 4. Conclusions

The signal time delay in the vicinity of the sun for the Dvali-Gabadadze-Porrati metric (DGP) has been calculated, for various angles in the range $30° \leq \theta \leq 72^0$. Both algebraic signs have been considered in the metric element. In particular the plus sign is related to the Friedman-Lemaitre-Robertson-Walker phase of the universe, and results to a positive time delay or an addition in the total traveling time due to Dvali-Gabadadze-Porrati, where the negative one is related to a self accelerating phase of the universe, results to a negative time delay or reduction in the total time due the contribution of DGP gravity correction. The signal delays are calculated in fractions of picoseconds. Signal delays of this magnitude might be in the borderline of time detection of today's technology and therefore there might be difficult to detect. Future technologies might be able to push for such a delectability limit, and therefore delays attribute to (DGP) gravity might be measured, in solar system experiments.